\documentclass{article}

\usepackage{microtype}
\usepackage{graphicx}
\usepackage{subfigure}

\usepackage{amssymb}
\usepackage{amsmath}
\usepackage{amsfonts} 
\usepackage{bm} 
\usepackage{booktabs} 

\usepackage{hyperref}

\usepackage[accepted]{icml2021}

\usepackage{multirow} 
\usepackage[table]{xcolor} 

\newcommand{\GongzhuSociety}{https://github.com/Gongzhu-Society}
\newcommand{\gongzhuonline}{http://cp.lovephy.cn:9000}

\begin{document}
\twocolumn[

\icmltitle{ScrofaZero: Mastering Trick-taking Poker Game Gongzhu by Deep Reinforcement Learning}



\icmlsetsymbol{equal}{*}

\begin{icmlauthorlist}
\icmlauthor{Naichen Shi}{equal,um}
\icmlauthor{Ruichen Li}{equal,pku}
\icmlauthor{Sun Youran}{equal,tsu}
\end{icmlauthorlist}

\icmlaffiliation{um}{IOE, University of Michigan}
\icmlaffiliation{pku}{EECS, Peking University}
\icmlaffiliation{tsu}{Yau Mathematical Sciences Center, Tsinghua University.}

\icmlcorrespondingauthor{Sun Youran}{syouran0508@gmail.com}

\icmlkeywords{Imperfect information, game, reinforcement learning}

\vskip 0.3in
]



\printAffiliationsAndNotice{\icmlEqualContribution} 


\begin{abstract}
People have made remarkable progress in game AIs,  especially in domain of perfect information game. However, trick-taking poker game, as a popular form of imperfect information game, has been regarded as a challenge for a long time. Since trick-taking game requires high level of not only reasoning, but also inference to excel, it can be a new milestone for imperfect information game AI. We study Gongzhu, a trick-taking game analogous to, but slightly simpler than contract bridge. Nonetheless, the strategies of Gongzhu are complex enough for both human and computer players. We train a strong Gongzhu AI ScrofaZero from \textit{tabula rasa} by deep reinforcement learning, while few previous efforts on solving trick-taking poker game utilize the representation power of neural networks. Also, we introduce new techniques for imperfect information game including stratified sampling, importance weighting, integral over equivalent class,  Bayesian inference, etc. Our AI can achieve human expert level performance. The methodologies in building our program can be easily transferred into a wide range of trick-taking games.
\end{abstract}
\section{Introduction}
We live in a world full of precariousness. Like a famous quotation from Sherlock Holmes, ``I have a turn both for observation and for deduction."\cite{holmes}, one should deduce hidden information from seemingly random observations to make good decisions. 
Imperfect information game is an abstraction of multi-agent decision making with private information. Related theory has been found useful in auction, mechanism design, etc \cite{gametheorytextbook}. The research of specific examples of imperfect information game can strengthen our abilities to navigate through the uncertainties of the world.

People have successfully built super-human AIs for perfect information games including chess \cite{chess} and Go \cite{alphagozero} by using deep reinforcement learning. Also, by combining deep learning with imperfect sampling, researchers have also made huge progress in Mahjong \cite{suphx}, StarCraft \cite{starcraft}, Dota \cite{dota}, and Texas hold'em \cite{texadholdem}. 

We study Gongzhu, a 4-player imperfect information poker game. Gongzhu is tightly connected with a wide range of trick-taking games. The detailed rules will be introduced in section \ref{sec::rules}. Building a strong Gongzhu program can deepen our understanding about imperfect information games.

We study Gongzhu for three reasons. Firstly, Gongzhu contains medium level of randomness and requires careful calculations to reign supreme. Compared with Mahjong and Texas hold'em, Gongzhu is more complicated since its decision space is larger, let alone toy poker games pervasively studied in like Leduc\cite{Leduc} and Kuhn\cite{Kuhn}. What's more, it is important to read the signals from the history of other players' actions and update beliefs about their private information continuously. The entanglement of sequential decision making and imperfect information makes it extremely nontrivial to find a good strategy out of high degree of noise.

Secondly, the scoring system of Gongzhu is relatively simple compared with bridge, since players don't bid in Gongzhu. Thus the reward function is easier to design, and we can focus on training highly skilled playing AI given such reward function. 

Thirdly, compared with large scale games like StarCraft or Dota, Gongzhu is more computationally manageable: all experiments in this paper can be done on only 2 Nvidia 2080Ti GPUs!

We train a Gongzhu program from \textit{tabula rasa} by self play without any prior human knowledge other than game rules. Our algorithm is a combination of Monte Carlo tree search and Bayesian inference. This method is an extension of MCTS to imperfect information game. Our program defeats expert level human Gongzhu players in our online platform \href{\gongzhuonline}{Gongzhu Online}.

We summarize our contributions below:
\begin{itemize}
\setlength{\itemsep}{0pt}
\item We introduce the game of Gongzhu that is more difficult than Leduc but more manageable than StarCraft. Gongzhu can be a benchmark for different multi-agent reinforcement learning algorithms.
\item We train a strong Gongzhu agent ScrofaZero purely by self-play. The training of ScrofaZero requires neither human expert data nor human guidance beyond game rules.
\item To our best knowledge, we are the first to combine Bayesian inferred importance sampling with deep neural network for solving trick-taking game. Our methods can be transferred to other trick-taking games including contract Bridge. 
\end{itemize}

The paper is organized as follows. In section \ref{sec::rules}, we review the rule of Gongzhu and its connection to other trick-taking games. In section \ref{sec::literature}, we discuss related work in literature. In section \ref{sec::framework}, we present an overview of our framework. In section \ref{sec::method}, we show some of our key methodologies including stratified sampling and integral over equivalent class. And in the section \ref{sec:empiricalanalysis}, we analyze the results of extensive empirical experiments.


\section{Rule of Gongzhu and Related Games}
\label{sec::rules}

Before delving further into our program, we will use a section to familiarize readers the rules of Gongzhu. We will introduce rules from general trick-taking poker game to specific game Gongzhu.

Gongzhu belongs to the class \textit{trick-taking}, which is a large set of games including bridge, Hearts, Gongzhu and Shengji. For the readers unfamiliar with any of these games, see supplementary material \ref{subsec:trick-taking-games} for the common rules of trick taking games. From now on, we assume readers understand the concept of \textit{trick}.

\begin{figure}[!htb]
    \centering
    \includegraphics[width=0.45\textwidth]{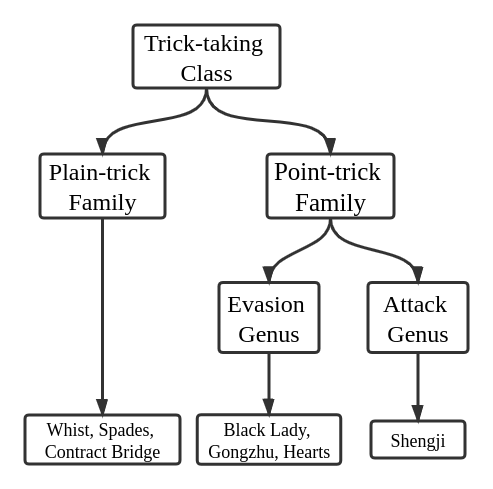}
    \caption{The classification of trick-taking games. Gongzhu belongs to the genus evasion, the family point-trick. As shown in this figure, Gongzhu is tightly connected with a wide variety of games.}
    \label{fig:trickgame}
\end{figure}

The class trick-taking is divided into two major families according to the goal of the games: family \textit{plain-trick} and family \textit{point-trick}. In family \textit{plain-trick}, like Whist, Contract bridge and Spades, the goal of games is to win specific or as many as possible  number of tricks. In family \textit{point-trick}, like Black Lady, Hearts, Gongzhu and Shengji, the goal of games is to maximize the total points of cards obtained.

For most of games in the family \textit{point-trick}, only some cards are associated with points. Depending on the point counting system, family \textit{point-trick} is furthermore subdivided into two genus, evasion and attack. For genus evasion, most of points are negative, so the strategy is usually to avoid winning tricks, while genus attack the opposite. Gongzhu belongs to the genus evasion. The points in Gongzhu are counted by following rules.
\begin{enumerate}
    \setlength{\itemsep}{0pt}
    \setlength{\parskip}{0pt}
    \setlength{\parsep}{0pt}
    \item Every heart is associated with points. Their points are shown in table \ref{tab:points}. Notice that,
    \begin{itemize}
        \item all points of hearts are non-positive;
        \item the higher rank a heart card has, the greater its point's absolute value will be;
        \item the total points of hearts are -200.
    \end{itemize}
    \item SQ has $-100$ points. SQ is called \textit{zhu} (scrofa) in this game. As \textit{zhu} has the most negative points, players will try their best avoid getting it.
    \item DJ has $+100$ points. DJ is called \textit{yang} (sheep/goat), in contrast with \textit{zhu}.
    \item C10 will double the points. However if one gets only C10, this C10 is counted as $+50$. C10 is called transformer for obvious reasons.
    \item If one player collects all the 13 hearts, to reward her braveness and skill, the points of all the hearts will be count as $+200$ rather than $-200$. This is called \textit{all hearts}. It is worth clarify that,
    \begin{itemize}
        \setlength{\itemsep}{0pt}
        \item to get \textit{all hearts}, one need get all 13 hearts, including zero point hearts H2, H3 and H4;
        \item the points are counted separately for each player and then summed together in each team.
    \end{itemize}
\end{enumerate}
All the rules except \textit{all hearts} are summarized in table \ref{tab:points}. The classification of trick-taking games and where Gongzhu belongs are shown in figure \ref{fig:trickgame}.

\begin{table}[!ht]
	\centering
    \begin{tabular}{|cc|cc|}
    \hline
    HA& $-50$& SQ& $-100$\\
    HK& $-40$& DJ& $+100$\\
    HQ& $-30$& & $+50$ or double\\
    HJ& $-20$& \multirow{-2}{*}{C10}&the points\\
    H5 to H10&$-10$&&\\
    H2 to H4&0&&\\
    \hline
    \end{tabular}
	\caption{Points of cards in Gongzhu. Gongzhu is a trick-taking class, point-trick family, evasion genus game. Note that except for points shown in the above table, there is an extra \textit{all hearts} rule explained in Section \ref{sec::rules}.}
	\label{tab:points}
\end{table}

\section{Literature review}
\label{sec::literature}

\subsection{Incomplete Information Game}
Different from Go, Gongzhu is a dynamic incomplete information game \cite{gametheorytextbook}. The research on the dynamics of imperfect information game and type of equilibrium has a long history, for example \cite{tremblinghand} introduces a more refined equilibrium concept called trembling hand equilibrium. 
Our work applies these theoretical analysis including Bayesian inference and sequential decision making to the trick-taking game Gongzhu. Recently, people proposed counterfactual regret minimization related algorithms \cite{regretminimization,discountedregretminimization} that can be proved to find an $\epsilon$-Nash equilibrium. The applications of counterfactual regret minimization type algorithms have been found successful in Texas hold'em \cite{texadholdem}. Such algorithms are not directly applicable to Gongzhu, which has a larger decision space and longer decision period.

\subsection{Computer Bridge}
\label{subsec:computerbridge}
As shown in section \ref{sec::rules}, Gongzhu is tightly connected with bridge. However unlike chess and go, computer bridge programs cannot beat human experts yet. The recent winners of the World Computer-Bridge Championship \cite{computerbridgewiki} are Wbridge5\footnote{\url{http://www.wbridge5.com/index.htm}} (on 2018) and Micro Bridge\footnote{\url{http://www.osk.3web.ne.jp/~mcbridge/index.html}} (on 2019)\footnote{The 2020 Championship is cancelled. The next championship will be in 2021.}. Micro Bridge first randomly generates unknown hands under known conditions derived from history, then apply tree search and pruning algorithms to make decision under perfect information. Wbridge5 does not reveal their algorithm to public, but it is believed to be similar to Micro Bridge, i.e. human-crafted rules for bidding and heuristic tree search algorithms for playing. Different from these work, we use deep neural network to evaluate current situation and to generate actions. Recently, \cite{bridgebidding} built a bidding program by supervised learning and then by reinforcement learning. In contrast, we train our Gongzhu program from \textit{tabula rasa}.

\subsection{Monte Carlo Tree search}
The use of Monte Carlo tree search both as a good rollout algorithm and a stable policy improvement operator is popular in perfect information game like go \cite{alphagozero}, and chess \cite{chess}. \cite{mctstheory} analyzed some theoretical properties of MCTS used in AlphaGo zero. \cite{ismcts,mctssurvey} discusses popular MCTS algorithms, including  information set Monte Carlo tree search (ISMCTS), an algorithm that combines MCTS with incomplete. The Monte Carlo tree search algorithm used in our program is the standard upper-confidence bound minimizing version, and computationally simpler than the full ISMCTS.

\subsection{Multi-agent RL}
The dynamics of training multi-player game AIs by self-play can be complicated. People have found counter-examples where almost all gradient-based algorithms fail to converge to Nash equilibrium \cite{impossibleconvergence}. Also, some games are nontransitive for certain strategies \cite{spinningtop}. Instead of a Nash equilibrium strategy, we attempt to find a strong strategy that can beat human expert level player. We also define a metric for the nontransitivity in the game. Different from \cite{sga}, where nontransitivity is defined only in the parameter space, our metric can be defined for any strategy. 

\section{Framework}
\label{sec::framework}
This section is an overview of our program. We start by introducing some notations. As discussed above, Gongzhu is a sequential game with incomplete information. The \textbf{players} are denoted as $\mathcal{N}=\{0,1,2,3\}$. We denote integer $u\in [0,52]$ as the \textbf{stage} of game, which can also be understand as how many cards have been played. We define \textbf{history} $h^u\in\mathcal{H}$ is a sequence of tuple $\{(i_t,a_{t})\}_{t=1,...u}$, where $t$ is stage, $i_t$ is the player to take action at stage $t$, and $a_{t}$ is the card her played. History represent as the history of card playing up to time $u$. We sometimes replace $h^u$ by $h$ for simplicity. We use $h(t)$ denote the $t$-th tuple in $h$. History $h$ is publicly observable to all players.  It's natural to assume that each player can perfectly recall, i.e. they can remember the history exactly. After dealing of each round, players will have their \textbf{initial hand cards} $c=\{c_i\}_{i=0,1,2,3}$. $c_i$ is private information for player i, and other players cannot see it. Also, we denote $c_i^u$ as the $i$-th player's remaining hand cards at stage $u$. 
Action $a\in\mathcal{A}$ is the card to play. We set $|\mathcal{A}|=52$. By the rule of Gongzhu, a player may not be able to play all cards depending on the suits of that trick, thus actual legal choice of $a$ might be smaller than the number of remaining cards in one's hand.

Information set is a concept in incomplete information game theory that roughly speaking, characterizes the information input for decision making. For standard Gongzhu game, we define the information set for player $i$ at time $u$ to be $(h^u,c_i)$, i.e. public information $h^u$ combined with player $i$'s private information $c_i$. The payoff $r$ is calculated only at the end of the game, i.e. when all 52 cards are played. $r(h^{T})=(r_0(h^{T}),r_1(h^{T}),r_2(h^{T}),r_3(h^{T}))$(where $T=52$) representing the scores for each player. A \textbf{strategy} of player $i$ is a map from history and private information to a probability distribution in action space: $\pi_i:\mathcal{H}\times\mathcal{C}_i\to\mathcal{A}$, i.e. given a specific history $h$ and an initial card $c_i$, $\pi_i(c_i,h)$ chooses a card to play. 
A \textbf{strategy profile} $\pi$ is the tuple of 4 players' strategies $\pi=\left(\pi_0,\pi_1,\pi_2,\pi_3\right)$. We use $\pi_{-i}$ to denote the strategy of other players except i. 

The value of an information set on the perspective of player $i$ is:
\begin{equation}
\begin{aligned}
\label{eqn::defofv}
v_i(\pi_i,\pi_{-i},c_i,h)&=\mathbb{E}_{p(c_{-i}|h,\pi)}\left[\mathbb{E}_{\pi}\left[r_i(h^{52})\right]\right]\\
&=\mathbb{E}_{p(c_{-i}|h,\pi)}\left[v^{\pi}(h^{52},c_i,c_{-i})\right]
\end{aligned}
\end{equation}
We use $v^{\pi}(h,c_i,c_{-i})$ replace $\mathbb{E}_{\pi}\left[r_i(h^{52})\right]$ for simplicity. $v^{\pi}(h,c_i,c_{-i})$ can be interpreted as value of a state under perfect information. $h^{52}$ is the history of all possible terminal state starting from $h$ with initial hands $c$. The inner expectation is taken in the following sense. Suppose there exists an oracle that knows exactly each players' initial cards c and each player's strategy $\pi$, it plays on each player's behalf with their strategy $\pi$ starting from $h$ till the end of the game. Due to the randomness of mixed strategy, the outcome, thus the payoff of the game is also random. The inner expectation is taken over the randomness of the game trajectory resulted from mixed strategy. 

The outer expectation is taken over player $i$'s belief. We define a \textbf{scenario} to be one possible initial hand configuration $c_{-i}$ from the perspective of player $i$. In Gongzhu, one player's belief is the probability $p(c|h,\pi)$ she assigns to each possible scenario. At the beginning of the game, all possible scenarios are of equal probability. This is the case when $h=h^0=\emptyset$, which is usually referred as common prior by game theorists. When the game proceeds, every player can see what have been played by other players, and this will change one's belief about the possibility of different scenarios. For example, clearly some initial hands configurations that are not compatible with the rule of Gongzhu will have probability 0. A natural way of updating belief is Bayes' rule. We will cover more details on calculating the outer expectation on section \ref{subsec:stratifiedsampling} and \ref{subsec:integral-over-EC}.

A player's goal is to optimize its strategy on every decision node, assuming that she knows other players' strategy $\pi_{-i}$:
\begin{equation}
\max_{\pi_i}v_i(\pi_i,\pi_{-i},c_i,h)
\end{equation}
She will choose an action on every decision node to maximize her expected payoff. Gongzhu ends after all 52 cards are played. The extensive form of the game can be regarded as a $13\times 4$-layer decision tree. Thus in principle, Bayes perfect equilibrium \cite{gametheorytextbook} can be solved by backward induction. However we do not to solve for exact Bayes perfect equilibrium here because (i) compared with obtaining a Bayes perfect equilibrium policy that is unexploitable by others, it's more useful to obtain a policy that can beat most other high level players. (ii) an exact Bayes perfect equilibrium is computationally infeasible.  

To obtain a strong policy, we train a neural network by self-play. Our neural network has a fully connected structure (see supplementary material \ref{subsec:networkarchitecture}). We divide training and testing into two separate processes. In training, we train this neural network to excel under perfect information, and in testing, we try to replicate the performance of model under perfect information by adding Bayesian inference described in section \ref{subsec:stratifiedsampling} and \ref{subsec:integral-over-EC}.

To train the neural network, we assume each player knows not only his or her initial hands, but also the initial hands of other players (see supplementary material \ref{subsec:perfectinfo}). In the terminology of computer bridge, this is called \textit{double dummy}. Then the outer expectation in equation \eqref{eqn::defofv} can be removed since each player knows exactly what other players have in their hands at any time of the game. The use of perfect information in training has two benefits, firstly the randomness of hidden information is eliminated thus the training of neural network becomes more stable, secondly since sampling hidden information is time consuming, using perfect information can save time. Although this treatment may downplay the use of strategies like bluffing in actual playing, the trained network performs well. Inspired by AlphaGo Zero \cite{alphagozero}, we use Monte Carlo tree search as a policy improvement operator to train the neural network (see supplementary material \ref{subsec:mcts}). The loss function is defined as:
\begin{equation}
\label{eqn::defloss}
\ell = \textrm{Div}_{\textrm{KL}}\left(p_{\textrm{MCTS}}||p_{\textrm{nn}}\right)+\lambda \left|v_{\textrm{MCTS}}-v_{\textrm{nn}}\right|.
\end{equation}
where $p_{\textrm{MCTS}}$ and $v_{\textrm{MCTS}}$ is the policy and value of a node returned by Monte Carlo tree search from that node, and $p_{\textrm{nn}}$ and $v_{\textrm{nn}}$ are the output of our neural network. The parameter $\lambda$ here weights between value loss and policy loss. Since the value of a specific node can sometimes be as large as several hundred while KL divergence is at most 2 or 3, this parameter is necessary. For more details in training, see supplementary material \ref{subsec:training}.

\begin{figure}[!ht]
    \centering
    \includegraphics[width=0.45\textwidth]{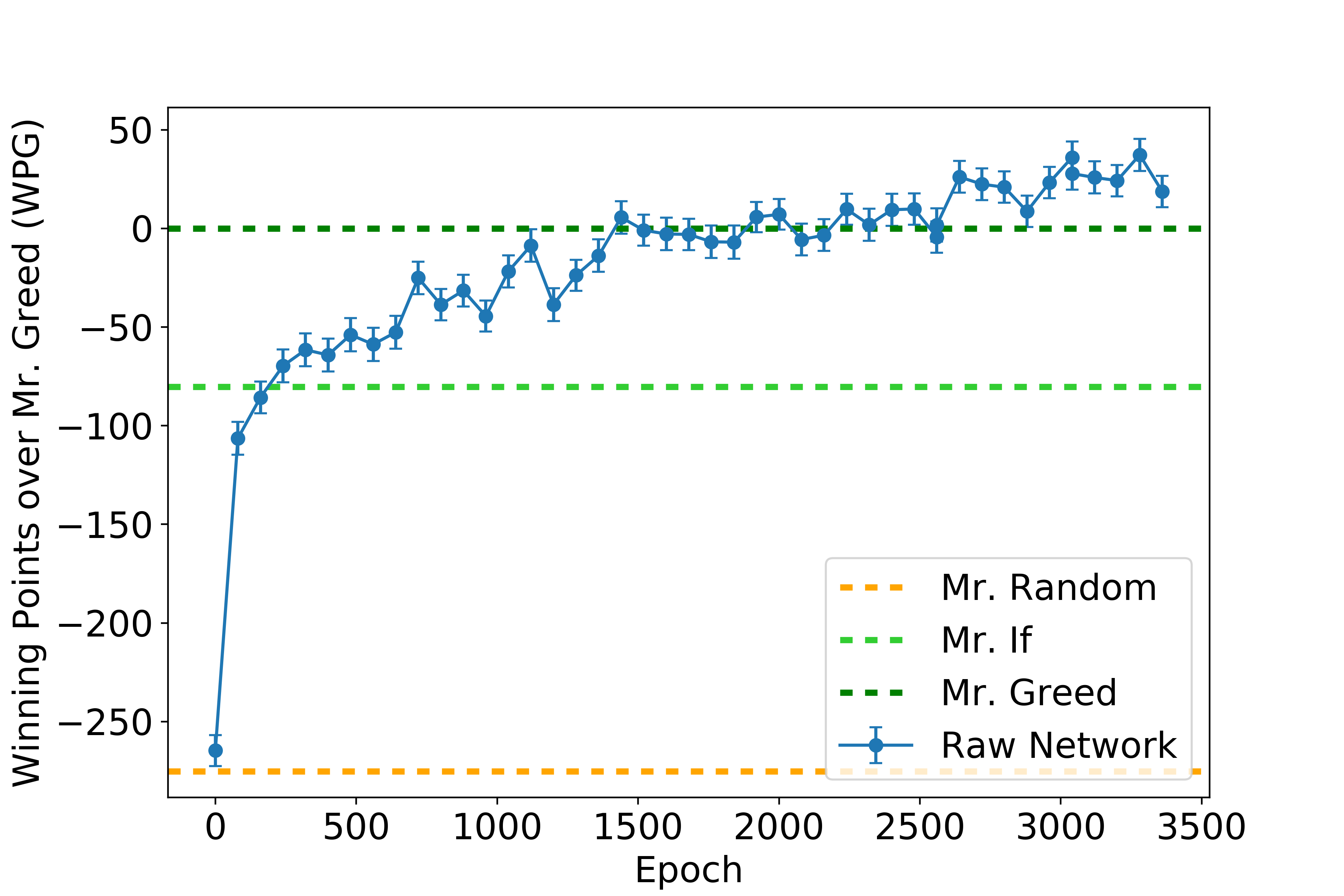}
    \caption{AI trained with perfect information and tested in standard Gongzhu game. Testing scores are calculated by WPG described in section \ref{subsec:eval_system}. Raw network means that we use MCTS for one step, with the value network as default evaluater. Mr. Random, If and Greed are three human experience AIs described in section \ref{subsec:classicalAI}. Every epoch takes less than one minute on a single Nvidia 2080Ti GPU.}
    \label{fig:best_result}
\end{figure}

For actual testing or benchmarking, however, we must comply with rules and play with honesty. On the other hand, the use of Monte Carlo tree search requires each player's hand to allow each player perform tree search. To bridge the gap we use stratified and importance sampling to estimate the outer expectation of equation \eqref{eqn::defofv}. We sample N scenarios by our stratified sampling, then use MCTS with our policy network as default value estimator to calculate the Q value for each choice. Then we average these Q values with an importance weight of hidden information. Finally, we choose the cards with the highest averaged Q value to play. Details of 
stratified and importance sampling will be discussed in section \ref{subsec:stratifiedsampling} and \ref{subsec:integral-over-EC}. In figure \ref{fig:best_result}, we can see that the neural network can improve itself steadily. 

\section{Methodology}
\label{sec::method}

\subsection{Human Experience Based AIs}
\label{subsec:classicalAI}
To expand the strategy space, we build a group of human experience based AIs using standard methods. We name them Mr. Random, Mr. If, and Mr. Greed. Among them, Mr. Greed is the strongest. The performance of these AIs are shown in table \ref{tab:transitivity}. More details can be found in supplementary material \ref{subsec:classicalAI-detail}.

\begin{table}[!ht]
    \centering
    \begin{tabular}{c|ccccc}
    \hline
&R&I&G&SZS\\
\hline
  R&    0&  194&  275&  319\\
  I& -194&    0&   80&  127\\
  G& -275&  -80&    0&   60\\
SZS& -319& -127&  -60&    0\\
\hline
    \end{tabular}
    \caption{Combating results for 4 different AIs. In this table, R stands for Mr. Random, I for Mr. If, G for Mr. Greed, SZS for ScrofaZeroSimple. R, I and G are human experience AIs described in Section \ref{subsec:classicalAI} and supplementary material \ref{subsec:classicalAI-detail}. SZS is ScrofaZero without IEC algorithm described in section \ref{subsec:integral-over-EC}.}
    \label{tab:transitivity}
\end{table}

\subsection{Evaluation System}
\label{subsec:eval_system}
We evaluate the performance of our AIs by letting two copies of AI team up as partners and play against two Mr. Greeds, and calculate their average winning points. We call this Winning Point over Mr. Greed (WPG). In typical evaluation, we run $1024\times 2$ of games. We show that this evaluation system is well-defined in supplementary material \ref{subsec:detail_eval_system}.


\subsection{Stratified Sampling}
\label{subsec:stratifiedsampling}
Given a history $h$, initial cards on player $i$'s hand $c_i$, one should estimate what cards other players have in their hands. As discussed before, we denote it by $c_{-i}$ and use $c_{-i}$ and scenario interchangeably. We use $\mathcal{C}(c_i)$ to denote the set of all possible $c_{-i}$'s. 

The most natural way to calculate one's belief about the distribution of scenarios in $\mathcal{C}(c_i)$ is Bayesian inference \cite{gametheorytextbook}. From Bayesian viewpoint, if player's strategy profile is $\pi_{i=0,1,2,3}$, which we now assume to be common knowledge, player $i$'s belief after observing history $h$ that initial cards in other players' hands are $c_{-i}$ is
\begin{equation}
\label{eqn::defofposteriordis}
p(c_{-i}|h)=\frac{p(h|c_{-i};\pi)p(c_{-i})}{\sum_{e\in \mathcal{C}(c_i)}p(h|e;\pi)p(e)}    
\end{equation}
where $p(h|c_{-i};\pi)$ is the probability that history $h$ is generated if initial card configuration is $c_{-i}$, and players play according to strategy profile $\pi_{i=0,1,2,3}$. We omit the dependence on $\pi$ when there is no confusion. 

The belief is important because players use it to calculate the outer expectation in equation \eqref{eqn::defofv}
\begin{equation}
\label{eqn::objectivedef}
\mathbb{E}_{c_{-i}\sim p(c_{-i}|h)}\left[v^{\pi}(h,c_i,c_{-i})\right]
\end{equation}
where $v^{\pi}(h,c_i,c_{-i})$ is the value function. The exact calculation of \eqref{eqn::objectivedef} through \eqref{eqn::defofposteriordis} requires calculating all possible configurations in set $\mathcal{C}(c_i)$, which can contain $39!\sim10^{46}$ elements. Such size is computationally intractable, we therefore seek to estimate it by Monte Carlo sampling. The naive application of Monte Carlo sampling can bring large variance to the estimation, thus we will derive a stratified importance sampling approach to obtain high quality samples.

For trick-taking games, there are always cases where some key cards are much more important than the others. 
These cards are usually associated with high variance in Q value, see section \ref{subsec:comparewithG}. Especially true is this for Gongzhu. 
We use stratified sampling to exhaust all possible configuration of important cards.

More specifically, we firstly divide the entire $\mathcal{C}(c_i)$ into several mutually exclusive strata $\{S_1,S_2,...S_p\}$, such that $\mathcal{C}(c_i)=\bigcup_{j=1}^tS_j$. Each stratum represents one important cards configuration. To generate the partition $\{S_1,S_2,...S_t\}$, we identify the key cards $\{c^k_1,c^k_2,...c^k_q\}$ in $c_{-i}$ based on the statistics of trained neural network, see section \ref{subsec:comparewithG}, then exhaust all possible configurations of $\{c^k_1,c^k_2,...c^k_q\}$. After obtaining the partition, we sample inside each stratum. More formally, by conditional expectation rule, we can rewrite equation \eqref{eqn::objectivedef} as
\begin{equation}
\label{eqn::stratifiedsampling}
    \sum_{j=1}^tp(S_j)\mathbb{E}_{c_{-i}\sim p(c_{-i}|h,S_j)}\left[v^{\pi}(h,c_i,c_{-i})\right]
\end{equation}
where $p(S_j)$ is the probability that $c_{-i}$ is in stratum $S_j$, $p(S_j)=\mathbb{E}_{c_{-i}\sim p(c_{-i}|h)}\left[\mathbf{1}_{c_{-i}\in S_j}\right]$, and $p(c_{-i}|h,S_j)$ is the probability distribution of $c_{-i}$ given the history $h$ and the fact that $c_{-i}$ is in stratum $S_j$. As a zero-th order approximation, we set $p(S_j)=\frac{1}{t}$ for all j.


Since the expectation in equation \eqref{eqn::stratifiedsampling} is still analytically intractable, we employ importance sampling to bypass the problem. If we can obtain a sample from a simpler distribution $q(c_{-i})$, which has common support with $p(c_{-i}|h)$, then by Radon-Nikodym theorem:
\begin{equation}
\label{eqn::importancesampling}
\begin{aligned}
&\mathbb{E}_{c_{-i}\sim p(c_{-i}|h)}\left[v^{\pi}(h,c_i,c_{-i})\right]\\
&=\mathbb{E}_{c_{-i}\sim q(c_{-i})}\left[v^{\pi}(h,c_i,c_{-i})\frac{p(c_{-i}|h)}{q(c_{-i})}\right]
\end{aligned}
\end{equation}
where we call the term $\frac{p(c_{-i}|h)}{q(c_{-i})}$ posterior distribution correction. If we draw N samples   $\mathcal{C}_N=\{c_{-i}^{(1)},c_{-i}^{(2)},...c_{-i}^{(N)}\}$ from $q(c_{-i})$:
\begin{equation}
\begin{aligned}
&\mathbb{E}_{c_{-i}\sim p(c_{-i}|h)}\left[v^{\pi}(h,c_i,c_{-i})\right]\\
&\approx\frac{1}{N}\sum_{k=1}^{N}v^{\pi}(h,c_i,c_{-i}^{(k)})\frac{p(c_{-i}^{(k)}|h)}{q(c_{-i}^{(k)})}
\end{aligned}
\end{equation}
We take $q(c_{-i})$ to be the following distribution:
\begin{equation}
q(c_{-i})=
\begin{cases}
  1/|\mathcal{C}(c_{-i})|&\text{if }c_{-i}\text{ is compatible with history}\\
  0&\text{otherwise}
\end{cases}
\end{equation}
i.e. $q(c_{-i})$ is a uniform distribution for all $c_{-i}$ that is compatible with history. Compatible with history means under such configuration actions in history h do not violate any rules.

Since the ratio $\frac{p(c_{-i}|h)}{q(c_{-i})}$ is still intractable, we use
\begin{equation}
\label{eqn::empprob}
    \hat{p}(c_{-i}^{(k)}|h)=\frac{p(h|c_{-i}^{(k)})p(c_{-i}^{(k)})}{\sum_{l=1}^Np(h|c_{-i}^{(l)})p(c_{-i}^{(l)})},\quad
    \hat{q}(c_{-i})=\frac{1}{N}
\end{equation}
to approximate $p(c_{-i}^{(k)}|h)$ and $q(c_{-i}^{(k)})$. Equation \eqref{eqn::empprob} change the the scope of summation on the denominator of  \eqref{eqn::defofposteriordis} from the entire population to only samples.
Then equation \eqref{eqn::importancesampling} reduces to:
\begin{equation}
\begin{aligned}
&\mathbb{E}_{c_{-i}\sim p(c_{-i}|h)}\left[v^{\pi}(h,c_i,c_{-i})\right]\\
&\approx \frac{1}{\sum_{k=1}^Ns(c_{-i}^{(k)})}\sum_{l=1}^{N}v^{\pi}(h,c_i,c_{-i}^{(l)})s(c_{-i}^{(l)})
\end{aligned}
\end{equation}
where $s(c_{-i}^{(k)})$ is the score we assign to scenario $c_{-i}^{(k)}$, it is defined as $s(c_{-i}^{(k)})=p(h|c_{-i}^{(k)})p(c_{-i}^{(k)})$. We will introduce an algorithm to calculate the score in section \ref{subsec:integral-over-EC}.

\subsection{Integral over Equivalent Class}
\label{subsec:integral-over-EC}
In this section, we will focus on how to compute $s(c_{-i})$. We assume that other players are using similar strategies to player $i$. Then the policy network of ScrofaZero can be used to estimate $p(h|c_i)$. To continue, we define \textit{correction factor} $\gamma$ for a single action as
\begin{equation}\label{eqn:defofcorfactor}
    \gamma (a,h,c_j)=e^{-\beta\cdot\textrm{regret}}=e^{-\beta(q_{\textrm{max}}-q_a)},
\end{equation}
to be the unnormalized probability of player $j$ taking action $a$ under the assumption of $j$ using similar strategies to player $i$. In definition \eqref{eqn:defofcorfactor}, $h$ is the history before action $a$, $c_j$ the hand cards for player $j$, $q_a$ the policy network output for player $j$'s action $a$ and $q_{\textrm{max}}$ the greatest value in outputs of legal choices in $c_j$, $\beta$ a temperature controlling level of certainty of our belief. Then the $p(h|c_{-i})$ in formula \eqref{eqn::defofposteriordis} can be written as
\begin{equation}\label{eqn:p_with_r}
\begin{aligned}
 p(h|c_{-i})&=p(h^u|c_{-i})=\prod_{t=0}^{u-1}p(a_{t+1}|h^{t},c_{j(t+1)})\\
 &=\prod_{t=0}^{u-1}\frac{\gamma(a_{t+1},h^t,c_{j(t+1)})}{\sum_{\alpha\in\textrm{lc}(t)} \gamma(\alpha,h^t,c_{j(t+1)})}~,
\end{aligned}
\end{equation}
where $\textrm{lc}(t)$ is legal choices at stage $t$. As a generalized approach of Bayesian treatment, we estimate $p(h|c_{-i})$ with products of \textit{correction factors}. We call this algorithm \textit{Integral over Equivalent Class} (IEC). The pseudocode for IEC is as algorithm \ref{alg:IEC}. As an attention mechanism and to save computation resources, some ``important" history slices are selected based on statistics in section \ref{sec:empiricalanalysis} in calculating the scenario's score in algorithm \ref{alg:IEC}, see supplementary material \ref{subsec:impotant-history-IEC} for detail.

Compared with naive Bayes weighting, our IEC weighting is insensitive to variations of number of legal choices thus more stable. Experiments show that IEC can outperform naive Bayes weighting by a lot, see table \ref{tab:rect-perform}.

\begin{algorithm}[!htb]
   \caption{Integral over Equivalent Class (IEC)}
   \label{alg:IEC}
\begin{algorithmic}
   \STATE {\bfseries Input:} history $h^u$, player $i$'s initial card $c_i$, one possible scenario $c_{-i}$.
   \STATE $s(c_{-i})\leftarrow 1$
   \FOR{$t = u-1,u-2,...,0$}
    \STATE $h \leftarrow h^t, c\leftarrow c_{j(t+1)}, a\leftarrow a_{t+1}$
    \IF{$a$ is important}
    \STATE $s(c_{-i})\leftarrow \gamma(a,h,c) s(c_{-i})$
    \ENDIF
   \ENDFOR
   \STATE {\bfseries Output:} Score for scenario $s(c_{-i})$.
\end{algorithmic}
\end{algorithm}

\begin{table}[ht]
	\centering
	{\rowcolors{2}{green!80!yellow!50}{green!60!yellow!30}
    \begin{tabular}{ccc}
    \hline
    Techniques& Performance& Win(+Draw) Rate\\
    \hline
    US& $59.6\pm3.3$& $62.2(+2.8)\%$\\
    SS& $59.7\pm3.3$& $64.4(+2.0)\%$\\
    US with IEC& $54.5\pm4.7$& -\\
    SS with IEC& $73.9\pm3.3$& $67.7(+2.5)\%$\\
    SS with IEC& &\\
    \rowcolor{green!60!yellow!30} (against US)&\multirow{-2}{*}{$15.5\pm4.1$}& \multirow{-2}{*}{$53.5(+3.3)\%$}\\
    \hline
	\end{tabular}}
	\caption{Performance after different methods. US stands for Uniformly Sampling, SS for Stratified Sampling, IEC for \textit{Integral over Equivalent Class}. The sampling number of US is set to 9 such that the sampling number will equal to that of SS. The last line of this table is ScrofaZero with the strongest sampling technique, SS with IEC, against itself without any method.}
	\label{tab:rect-perform}
\end{table}

In rest of this section we will explain the intuition of \textit{integral over equivalent class}. We begin by introducing the concept of \textit{irrelevant cards}. \textit{Irrelevant cards} should be the cards which (i) will not change both its correction factor and other cards' correction factors if it is moved to others' hand, and (ii) will not change its correction factor if other cards are moved. The existence of approximate \textit{irrelevant cards} can be confirmed both from experiences of playing games or from the statistics in section \ref{sec:empiricalanalysis}. In figure \ref{fig:val_var} of section \ref{subsec:empiricalanalysis-1}, we see that there are some cards whose variance of values are small. These cards are candidates approximate \textit{irrelevant cards}. See supplementary material \ref{subsec:example-irrelevant-cards} for an concrete example. 

We call two distributions of cards only different in irrelevant cards equivalent. This equivalent relation divides all scenarios $\mathcal{C}(c_{-i})$ into equivalent classes. We denote the equivalent class of scenario $c_{-i}$ as $\left[c_{-i}\right]$. We should integrate over the whole equivalent class once we get the result of one represent element, because the MCTS procedure for each scenario is expensive. The weight of one equivalent class should be
\begin{equation}\label{eqn:int_p}
    p(h^u|c_{-i})p(\left[c_{-i}\right])=\sum_{\substack{\textrm{all permutations of}\\\textrm{irrelevant cards}}}\prod_{t=0}^{u-1}\frac{\gamma(a_{t+1},h^t,c_{j(t+1)})}{Y_{t+1}+J_{j(t+1)}}
\end{equation}
where $j(t)$ is the player who played at stage $t$, $J_{j(t)}$ the sum of correction factors of irrelevant cards in $j(t)$ and $Y_t$ the sum of correction factors of other cards in $j(t)$. $Y$ may change in different scenarios but $\sum_{j=1}^3 J_j$ will keep unchanged by definition, denoted by $J$.

Follow the steps in supplementary material \ref{subsec:detail-IEC}, we can obtain the result of the summation in equation \eqref{eqn:int_p}
\begin{equation}\label{eqn:int_p_result}
p(h^u|c_{-i})p(\left[c_{-i}\right]) = 3^N \prod_{t=0}^{u-1}\frac{\gamma(a_{t+1},h^t,c_{j(t+1)})}{Y_{t+1}+J/3}+O(\xi^3)
\end{equation}
where $N$ is the number of irrelevant cards, $J$ the sum of correction factor of all irrelevant cards and $\xi$ a real number between $0$ and $2/3$. One can see the supplementary material \ref{subsec:detail-IEC} for detail.

But notice that, the denominators in the result of \eqref{eqn:int_p_result} are insensitive to change of $Y$ because both $Y$ and $J$ are always greater than 1 (see section \ref{subsec:empiricalanalysis-1} for the magnitude of $Y$ and $J$). For scenarios in different equivalent classes, their $Y$'s might be different but the $J$ is always the same. So the integral remains approximately the same. Thus we can ignore the denominators when calculating scenario's score. Or in other words, we can use the product of unnormalized correction factors as scenario's score. This is exactly the procedure of IEC. 

\section{Empirical Analysis}
\label{sec:empiricalanalysis}

\subsection{Neural Network Output Statistics}
\label{subsec:empiricalanalysis-1}
\begin{figure}[!h]
    \centering
    \includegraphics[width=0.45\textwidth]{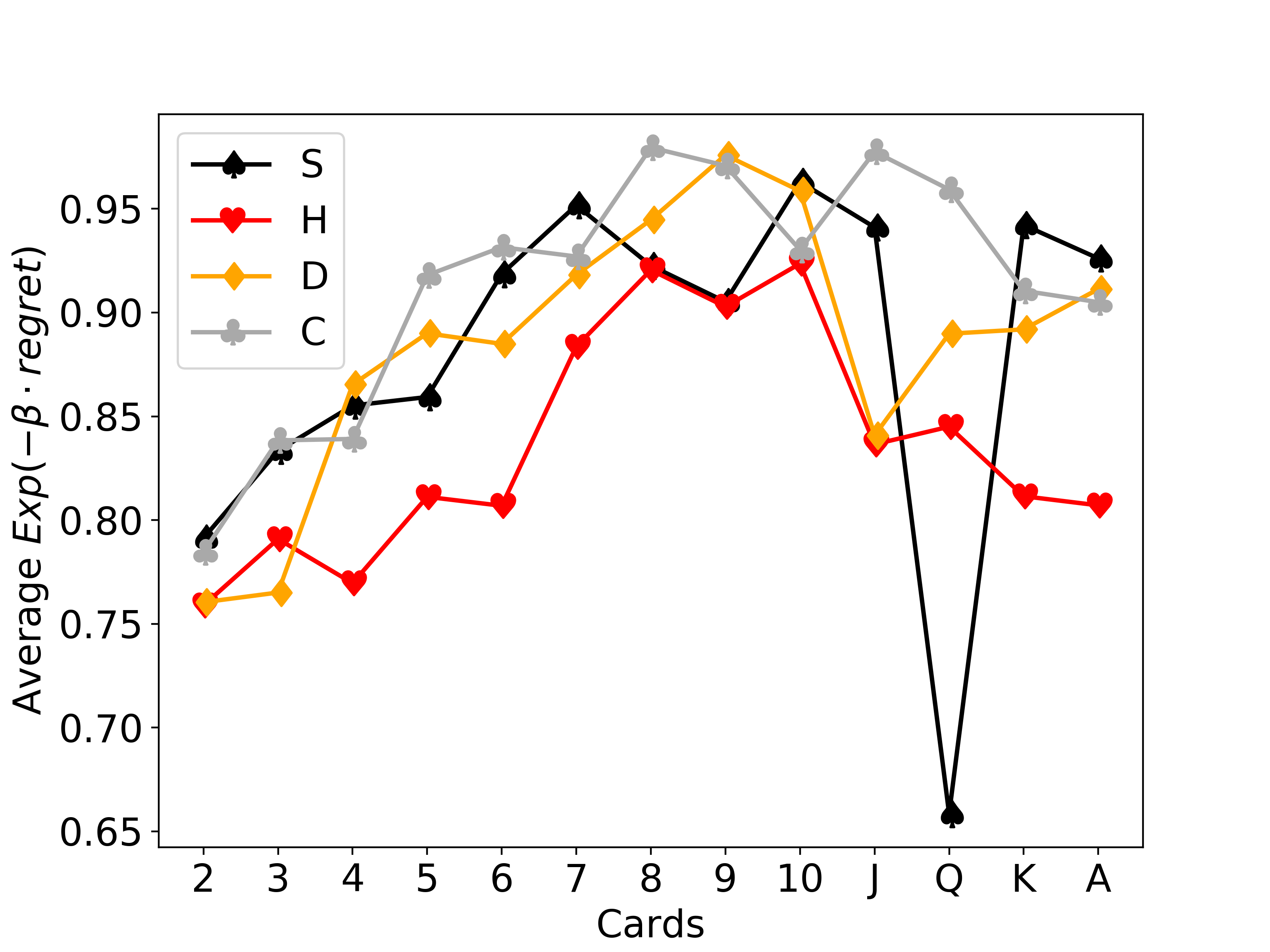}
    \caption{Average value of correction factor for different cards. The $\beta$ used here is equal to $0.015$.}
    \label{fig:reg}
\end{figure}

To begin with, we present some basic statistics for neural network. They include mean and variance of value and correction factor $\gamma$ of playing a card. The average correction factor $\gamma$ shown in figure \ref{fig:reg} reflect cards are ``good" or not: the higher correction factor, the better to take such action. We can find in the figure that, SQ is not a ``good" card, one would better not play it. Another finding is that the correction factor of four suits peak at around 10, thus when one player choose between cards lower than 10, she should play the largest one.  

\begin{figure}[!ht]
    \centering
    \includegraphics[width=0.45\textwidth]{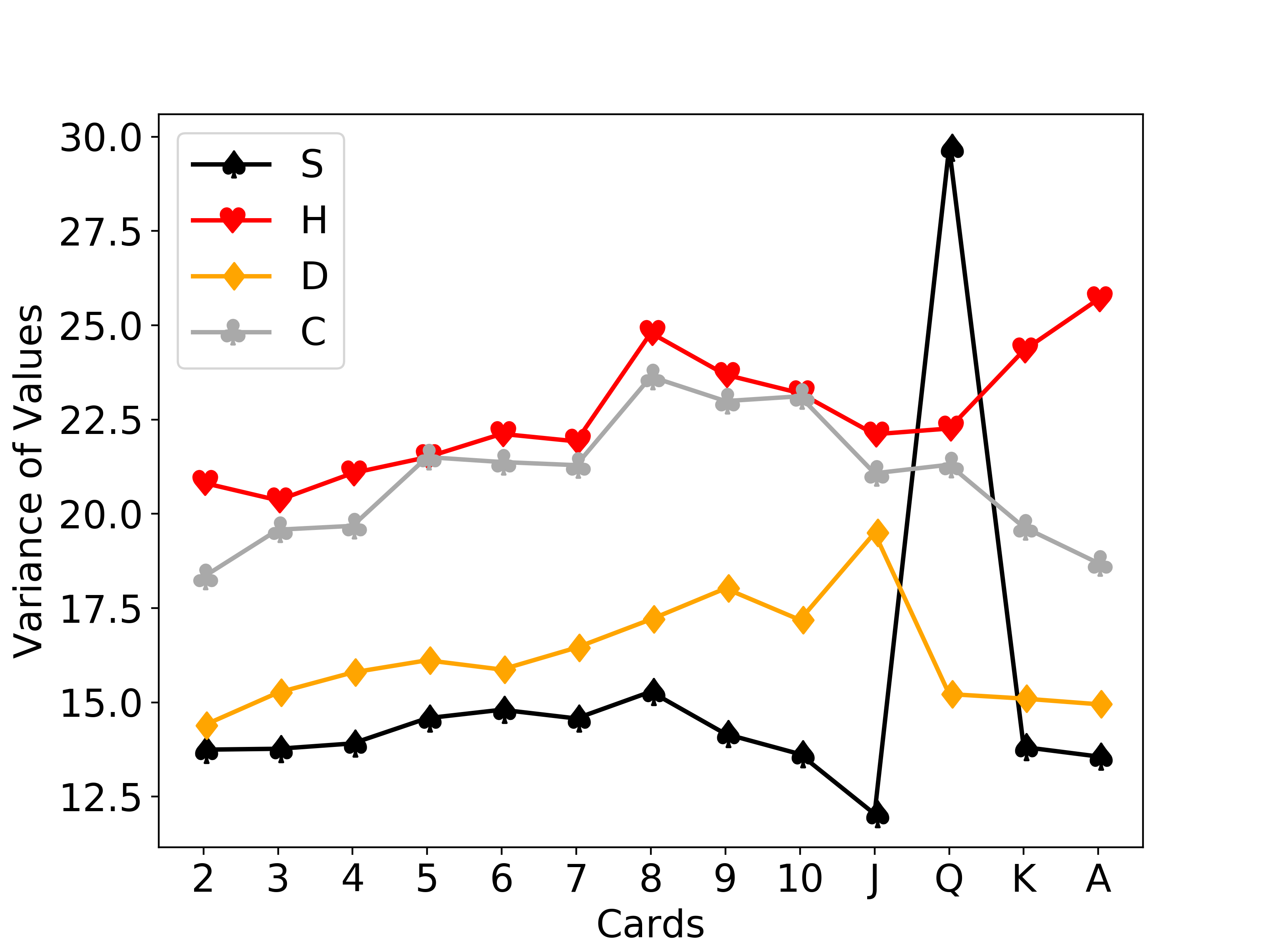}
    \caption{Variance of values for different cards.}
    \label{fig:val_var}
\end{figure}

However, for Gongzhu, the value of a card highly depends on the situation. Hence it's important to study the variance of correction factor. For example, a SQ will bring large negative points ends in your team but will bring large profits ends in your opponent. Variance of values shown in figure \ref{fig:val_var} illustrates the magnitude of risk when dealing with corresponding cards. We can see that SQ's variance is large, which is in line with our analysis. Meanwhile, heart cards should be dealt differently under different situations. Among heart suit, HK and HA are especially important. This may be the result of \textit{finesse} technique and \textit{all hearts} in Gongzhu.

These statistics from ScrofaZero reveal which cards are important. This information is used in stratified sampling in section \ref{subsec:stratifiedsampling}. Also, they are consistent with human experience.

\subsection{Comparison between Neural Network and Human Experience}
\label{subsec:comparewithG}
The best classical AI Mr. Greed explained in Section \ref{subsec:classicalAI} has many fine-tuned parameters, including the value for each card. For example, although SA and SK are not directly associated with points in the rule of Gongzhu, they have a great chance to get the most negative card SQ which weights $-100$. So SA and SK are counted as $-50$ and $-30$ points respectively in Mr. Greed. These parameters are fine-tuned by human experience. In the area of chess, people have compared the difference for chess piece relative value between human convention and deep neural network AI trained from zero \cite{chess_piece_points}. Here we will conduct a similar analysis to our neural network AI and Mr. Greed's parameters. Table \ref{tab:comparewithG} shows experience parameters in Mr. Greed for some important cards and ScrofaZero's output under typical situations. Negative means that card is a risk or hazard, thus it is better to get rid of it, while positive value has the opposite meaning. We can see that Mr. Greed and ScrofaZero agrees with each other very well.

\begin{table}[!ht]
    \centering
    {\rowcolors{2}{green!80!yellow!50}{green!60!yellow!30}
    \begin{tabular}{c|c|c}
         Cards& Mr. Greed& ScrofaZero\\
         \hline
         SA& $-50$& $-20\sim-40$\\
         SK& $-30$& $-20\sim-40$\\
         CA& $-20$& $-20$\\
         CK& $-15$& $-20$\\
         CQ& $-10$& $-10$\\
         CJ& $-5$& $-10$\\
         DA& $30$& $20$\\
         DK& $20$& $10$\\
         DQ& $10$& $10$\\
    \end{tabular}}
    \caption{The experience parameters in Mr. Greed and output of ScrofaZero. Negative means that card is a burden, positive the opposite. ScrofaZero's values are estimated under typical situations.}
    \label{tab:comparewithG}
\end{table}

\section{Conclusion}
In this work we introduce trick-taking game Gongzhu as a new benchmark for incomplete information game. We train ScrofaZero, a human expert level AI capable of distilling information and updating belief from history observations. The training starts from \textit{tabula rasa} and does not need domain of human knowledge. We introduce stratified sampling and IEC to boost the performance of ScrofaZero. 

Future research directions may include designing better sampling techniques, incorporating sampling into neural network, and applying our methods to other trick-taking games like contract bridge. Also, we believe the knowledge in training ScrofaZero can be transferred to other real world applications where imperfect information plays a key role for decision making. 


\bibliography{ref.bib}

\begin{thebibliography}{23}
\providecommand{\natexlab}[1]{#1}
\providecommand{\url}[1]{\texttt{#1}}
\expandafter\ifx\csname urlstyle\endcsname\relax
  \providecommand{\doi}[1]{doi: #1}\else
  \providecommand{\doi}{doi: \begingroup \urlstyle{rm}\Url}\fi

\bibitem[Brown \& Sandholm(2019)Brown and
  Sandholm]{discountedregretminimization}
Brown, N. and Sandholm, T.
\newblock Solving imperfect-information games via discounted regret
  minimization.
\newblock In \emph{The Thirty-Third AAAI Conference on Artificial
  Intelligence}, 2019.

\bibitem[Brown et~al.(2019)Brown, , and Sandholm]{texadholdem}
Brown, N., , and Sandholm, T.
\newblock Superhuman ai for multiplayer poker.
\newblock \emph{Science}, 2019.

\bibitem[Browne et~al.(2012)Browne, Powley, Whitehouse, Lucas, Cowling,
  Rohlfshagen, Tavener, Perez, Samothrakis, and Colton]{mctssurvey}
Browne, C., Powley, E., Whitehouse, D., Lucas, S., Cowling, P.~I., Rohlfshagen,
  P., Tavener, S., Perez, D., Samothrakis, S., and Colton, S.
\newblock A survey of monte carlo tree search methods.
\newblock \emph{IEEE Transactions on Computational Intelligence and AI in
  Games}, 4, 2012.

\bibitem[Bubeck \& Cesa-Bianchi(2012)Bubeck and Cesa-Bianchi]{banditbounds}
Bubeck, S. and Cesa-Bianchi, N.
\newblock Regret analysis of stochastic and nonstochastic multi-armed bandit
  problems.
\newblock \emph{Foundations and Trends in Machine Learning}, 5, 2012.

\bibitem[Czarnecki et~al.(2020)Czarnecki, Gidel, Tracey, Tuyls, Omidshafiei,
  Balduzzi, and Jaderberg]{spinningtop}
Czarnecki, W.~M., Gidel, G., Tracey, B., Tuyls, K., Omidshafiei, S., Balduzzi,
  D., and Jaderberg, M.
\newblock Real world games look like spinning tops.
\newblock In \emph{NeurIPS}, 2020.

\bibitem[Doyle()]{holmes}
Doyle, A.~C.
\newblock \emph{The Sign of Four}.

\bibitem[Grill et~al.(2020)Grill, Altché, Tang, Hubert, Valko, Antonoglou, and
  Munos]{mctstheory}
Grill, J.-B., Altché, F., Tang, Y., Hubert, T., Valko, M., Antonoglou, I., and
  Munos, R.
\newblock Monte-carlo tree search as regularized policy optimization.
\newblock In \emph{International Conference on Machine Learning}, 2020.

\bibitem[Kuhn(1950)]{Kuhn}
Kuhn, H.~W.
\newblock Simplified two-person poker.
\newblock \emph{Contributions to the Theory of Games}, 1950.

\bibitem[Letcher(2021)]{impossibleconvergence}
Letcher, A.
\newblock On the impossibility of global convergence in multi-loss
  optimization.
\newblock In \emph{International Conference on Learning Representations}, 2021.
\newblock URL \url{https://openreview.net/forum?id=NQbnPjPYaG6}.

\bibitem[Letcher et~al.(2019)Letcher, Balduzzi, Racani`ere, Martens, Foerster,
  Tuyls, and Graepel]{sga}
Letcher, A., Balduzzi, D., Racani`ere, S., Martens, J., Foerster, J., Tuyls,
  K., and Graepel, T.
\newblock Differentiable game mechanics.
\newblock \emph{Journal of Machine Learning Research}, 2019.

\bibitem[Li et~al.(2020)Li, Koyamada, Ye, Liu, Wang, Yang, Zhao, Qin, Liu, and
  Hon]{suphx}
Li, J., Koyamada, S., Ye, Q., Liu, G., Wang, C., Yang, R., Zhao, L., Qin, T.,
  Liu, T.-Y., and Hon, H.-W.
\newblock Suphx: Mastering mahjong with deep reinforcement learning.
\newblock \emph{arXiv}, 2020.
\newblock URL \url{https://arxiv.org/abs/2003.13590}.

\bibitem[OpenAI et~al.(2019)OpenAI, Berner, Brockman, Chan, Cheung, Debiak,
  Dennison, Farhi, Fischer, Hashme, Hesse, Józefowicz, Gray, Olsson, Pachocki,
  Petrov, de~Oliveira~Pinto, Raiman, Salimans, Schlatter, Schneider, Sidor,
  Sutskever, Tang, Wolski, and Zhang]{dota}
OpenAI, Berner, C., Brockman, G., Chan, B., Cheung, V., Debiak, P., Dennison,
  C., Farhi, D., Fischer, Q., Hashme, S., Hesse, C., Józefowicz, R., Gray, S.,
  Olsson, C., Pachocki, J., Petrov, M., de~Oliveira~Pinto, H.~P., Raiman, J.,
  Salimans, T., Schlatter, J., Schneider, J., Sidor, S., Sutskever, I., Tang,
  J., Wolski, F., and Zhang, S.
\newblock Dota 2 with large scale deep reinforcement learning.
\newblock 2019.
\newblock URL \url{https://arxiv.org/abs/1912.06680}.

\bibitem[Rong et~al.(2019)Rong, Qin, and An]{bridgebidding}
Rong, J., Qin, T., and An, B.
\newblock Competitive bridge bidding with deep neural networks.
\newblock \emph{Proceedings of the 18th International Conference on Autonomous
  Agents and MultiAgent Systems}, 2019.

\bibitem[Selten(1975)]{tremblinghand}
Selten, R.
\newblock Reexamination of the perfectness concept for equilibrium points in
  extensive games.
\newblock \emph{International journal of game theory}, 1975.

\bibitem[Silver et~al.(2017)Silver, Schrittwieser, Simonyan, Antonoglou,
  Aja~Huang, Hubert, Baker, Lai, Bolton, Chen, Lillicrap, Hui, Sifre, van~den
  Driessche, Graepel, and Hassabis]{chess}
Silver, D., Schrittwieser, J., Simonyan, K., Antonoglou, I., Aja~Huang, A.~G.,
  Hubert, T., Baker, L., Lai, M., Bolton, A., Chen, Y., Lillicrap, T., Hui, F.,
  Sifre, L., van~den Driessche, G., Graepel, T., and Hassabis, D.
\newblock Mastering the game of go without human knowledge.
\newblock \emph{Nature}, 2017.

\bibitem[Silver et~al.(2018)Silver, Hubert, Schrittwieser, Antonoglou, Lai,
  Guez, Lanctot, Sifre, Kumaran, Graepel, Lillicrap, Simonyan, and
  Hassabis]{alphagozero}
Silver, D., Hubert, T., Schrittwieser, J., Antonoglou, I., Lai, M., Guez, A.,
  Lanctot, M., Sifre, L., Kumaran, D., Graepel, T., Lillicrap, T., Simonyan,
  K., and Hassabis, D.
\newblock A general reinforcement learning algorithm that masters chess, shogi,
  and go through self-play.
\newblock \emph{Science}, 2018.

\bibitem[Southey et~al.(2005)Southey, Bowling, Larson, Piccione, Burch,
  Billings, and Rayner]{Leduc}
Southey, F., Bowling, M.~P., Larson, B., Piccione, C., Burch, N., Billings, D.,
  and Rayner, C.
\newblock Bayes' bluff: Opponent modelling in poker.
\newblock \emph{Proceedings of the Twenty-First Conference on Uncertainty in
  Artificial Intelligence}, 2005.

\bibitem[Tadelis(2013)]{gametheorytextbook}
Tadelis, S.
\newblock \emph{Game Theory: An Introduction}.
\newblock Princeton University press, 41 William Street, Princeton, New Jersey,
  2013.

\bibitem[Tomašev et~al.(2020)Tomašev, Paquet, Hassabis, and
  Kramnik]{chess_piece_points}
Tomašev, N., Paquet, U., Hassabis, D., and Kramnik, V.
\newblock Assessing game balance with alphazero: Exploring alternative rule
  sets in chess, 2020.

\bibitem[Vinyals et~al.(2019)Vinyals, Babuschkin, Czarnecki, Mathieu, Dudzik,
  Chung, Choi, Powell, Ewalds, Georgiev, Oh, Horgan, Kroiss, Danihelka, Huang,
  Sifre, Cai, Agapiou, Jaderberg, Vezhnevets, Leblond, Pohlen, Dalibard,
  Budden, Sulsky, Molloy, Paine, Gulcehre, Wang, Pfaff, Wu, Ring, Yogatama,
  Wünsch, McKinney, Smith, Schaul, Lillicrap, Kavukcuoglu, Hassabis, and
  an~David~Silver]{starcraft}
Vinyals, O., Babuschkin, I., Czarnecki, W.~M., Mathieu, M., Dudzik, A., Chung,
  J., Choi, D.~H., Powell, R., Ewalds, T., Georgiev, P., Oh, J., Horgan, D.,
  Kroiss, M., Danihelka, I., Huang, A., Sifre, L., Cai, T., Agapiou, J.~P.,
  Jaderberg, M., Vezhnevets, A.~S., Leblond, R., Pohlen, T., Dalibard, V.,
  Budden, D., Sulsky, Y., Molloy, J., Paine, T.~L., Gulcehre, C., Wang, Z.,
  Pfaff, T., Wu, Y., Ring, R., Yogatama, D., Wünsch, D., McKinney, K., Smith,
  O., Schaul, T., Lillicrap, T., Kavukcuoglu, K., Hassabis, D., and
  an~David~Silver, C.~A.
\newblock Grandmaster level in starcraft ii using multi-agent reinforcement
  learning.
\newblock \emph{Nature}, 2019.

\bibitem[Whitehouse(2004)]{ismcts}
Whitehouse, D.
\newblock \emph{Monte Carlo Tree Search for games with Hidden Information and
  Uncertainty}.
\newblock PhD thesis, University of York, 7 2004.

\bibitem[Wikipedia(2021)]{computerbridgewiki}
Wikipedia.
\newblock Computer bridge, 2021.
\newblock URL \url{https://en.wikipedia.org/wiki/Computer_bridge}.

\bibitem[Zinkevich et~al.(2007)Zinkevich, Johanson, Bowling, and
  Piccione]{regretminimization}
Zinkevich, M., Johanson, M., Bowling, M., and Piccione, C.
\newblock Regret minimization in games with incomplete information.
\newblock In \emph{20th Conference on Neural Information Processing Systems},
  2007.

\end{thebibliography}
\bibliographystyle{icml2021}

\newpage
\clearpage
\section{Experimental Details and Extend Data}

\subsection{Trick-taking Games}
\label{subsec:trick-taking-games}
Gongzhu belongs to the class trick-taking, which is a large set of games including bridge, Hearts, Gongzhu and Shengji. We dedicate this section to familiarizing readers with trick-taking games. Trick-taking games share the following common rules.
\begin{enumerate}
    \item A standard 52-card deck is used in most cases.
    \item Generally, there are four players paired in partnership, with partners sitting opposite to each other around a table. 
    \item Cards are shuffled and dealt to four players at the beginning.
    \item As the name suggests, trick-taking game consists of a number of tricks. In a trick, four players play one card sequentially by the following rules:
    \begin{itemize}
        \item The player leading the first trick is chosen randomly or by turns. The first card of each trick can be any card in that player's hand.
        \item Following players should follow the suit if possible. There are no limits on the ranking of the cards played.
        \item At the end of each trick, four cards played are ranked and the player who played the card of highest rank becomes the winner.
        \item The winner of the last trick leads the next trick.
        \item The playing order is usually clockwise.
    \end{itemize}
    \item The cards are usually ranked by: A K Q J 10 9 8 7 6 5 4 3 2. 
\end{enumerate}

\subsection{Details of Human Experience Based AIs}
\label{subsec:classicalAI-detail}
We build a group of human experience based AIs using standard methods. The group includes
\begin{enumerate}
    \item Mr. Random: Random player choosing cards from legal choices randomly.
    \item Mr. If: A program with 33 if statements representing human experience. Mr. If can outperform Mr. Random a lot with such a few number of ifs.
    \item Mr. Greed: AI with hundreds of if statements and partial backward induction. It contains many handcrafted hyper-parameters to complete the search. Mr. Greed can outperform Mr. If, but not proportional to their number of if statements. 
\end{enumerate}
The performances of these AIs are shown in table \ref{tab:transitivity}. More details can be found in \href{\GongzhuSociety}{our repository}.

\subsection{Network Architecture}
\label{subsec:networkarchitecture}
We use fully connected layer with skip connection as our model basic block. The input of our neural network is encoded into a 434-dimensional onehot vector (this will be explained in detail in the next subsection). The output is a 53 dimensional vector with the first 52 elements to be $p$ vector and the last element to be $v$.

We also tried other network topologies, including shallower fully connection layers and ResNet. Their performance are shown in table \ref{tab:net-perform}.

\begin{table}[h!]
	\centering
	{\rowcolors{2}{green!80!yellow!50}{green!60!yellow!30}
    \begin{tabular}{ccc}
    \hline
    Network Topology&\#Parameters&Performance\\
    \hline
    Fully Connection 16& 9315381& $0\pm8$\\
    Fully Connection 24& 11416299& $32\pm5$\\
    ResNet 18& 11199093& $-66\pm8$\\
    \hline
    \end{tabular}}
	\caption{Performance of different networks. These scores are evaluated by WPG introduced in section \ref{subsec:eval_system}. Sufficient rounds of games are played to make sure the variance is small enough.}
	\label{tab:net-perform}
\end{table}
Since ResNet does not show significant improvement in performance, we stick to fully connected neural network for most of our experiments.

\subsection{Network Inputs}
\label{subsec:networkinputs}
The input of our neural network is a $52\times4+54\times3+16\times4=434$ dimension vector.
\begin{itemize}
    \item $52\times4$ denotes the hands of 4 players. $52$ because Gongzhu uses standard 52-card deck. Cards of the other 3 players are guessed by the methods discussed in Section \ref{subsec:stratifiedsampling}.
    \item $54\times3$ denotes the cards played in this trick. We choose these format because at most 3 cards are played before the specified player can play. We use $54$ instead of $52$ to represent the cards due to the \textit{diffuse technique} described in the next subsection.
    \item $16\times4$ denotes the cards associated with points that is already played. In the game Gongzhu, there are $16$ cards which have scores.
\end{itemize}

\subsection{Diffuse Technique in Inputs}
\label{subsec:diffuse}
\begin{figure}[!ht]
    \centering
    \includegraphics[width=.45\textwidth]{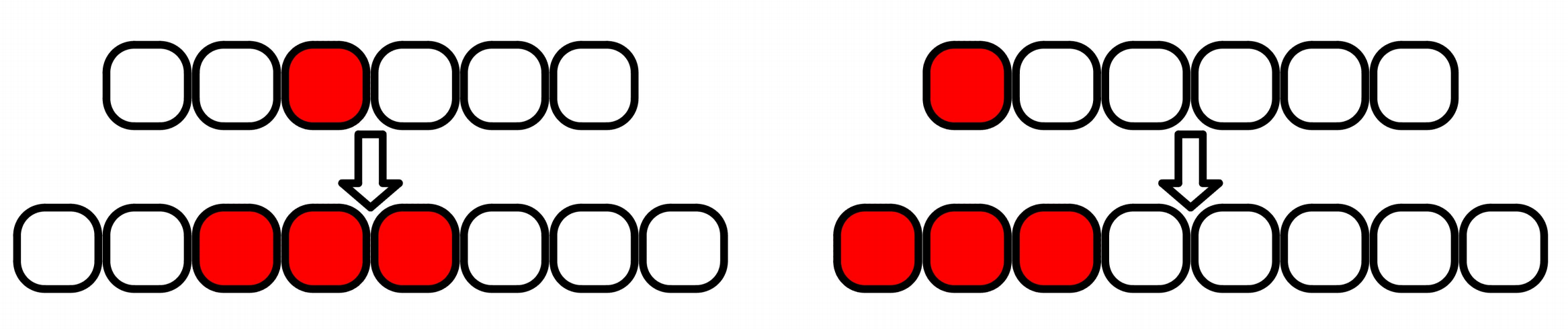}
    \caption{We use diffuse technique when representing cards in this trick to accelerate training. Above shows the normal input, while below shows the input after using diffuse technique.}
    \label{fig:diffuse}
\end{figure}

    We use \textit{diffuse technique} for representing cards in this trick when preparing inputs. Normally we should set one specific element of onehot vector corresponding to a card to $1$. However, we want to amplify the input signal. Hence we set not only the element in the onehot vector corresponding to this card, but also the two adjacent elements, to 1. Also we extend length for representing each card from $52$ and to $54$ to diffuse the element at two endpoints. Figure \ref{fig:diffuse} shows how diffuse technique works. Input in this form can be transformed to and from standard input with a single fully connection layer. In experiments, diffuse technique accelerates training.

\subsection{MCTS}
\label{subsec:mcts}
We used the standard MCTS using the value network to evaluate positions in the tree. We use UCB (Upper Confidence Bound)\cite{banditbounds} method to select a child node. More specifically, for each selection, we choose node $i=\mathop{\text{argmax}}_{j} v_j + c\sqrt{\ln N/n_j}$, where $v_j$ is the value of node $j$, $n_j$ is how many times node $j$ has been visited, $N$ is the total visit round, $c$ is the exploration constant. There are two hyperparameters important in MCTS: exploration constant $c$ and search number $T_{\textrm{MCTS}}$. We keep exploration constant to $c=30$. As the search number, we set $T_{\textrm{MCTS}}=2\times \{\textrm{legal choice number}\}$ when training and $T_{\textrm{MCTS}}=10+2\times\{\textrm{legal choice number}\}$ when evaluating. The search number is crucial in training and will be discussed in the next subsection.


\subsection{Collect Training Data under Perfect Information}
\label{subsec:perfectinfo}
The network is trained under perfect information where it can see cards in others' hands. Or in other words, we do not need to sample hidden information during training. This setting might be preferred because
\begin{enumerate}
    \item without sampling, training is more robust;
    \item by this means, more diversified middlegames and endings can be reached, which helps neural network improve faster by learning from different circumstances.
\end{enumerate}
For example, we find neural network trained with perfect information masters the techniques in \textit{all hearts} much more faster than one trained with imperfect information. 

After each MCTS search and play, the input and the rewards for different cards will be saved in buffer for training. The search number in MCTS is crucial. When it is small, the neural network of course will not improve. However, we find that, when search number in MCTS is too large, the neural network will again not improve, or even degrade! We find $2\times \{\textrm{legal choice number}\}$ the most suitable number for MCTS searching. Notice that, with $2\times \{\textrm{legal choice number}\}$ searches, MCTS can only predict the future approximately after two cards are played. It is quite surprising that the neural network can improve and finally acquire ``intuition" for long term future.

\subsection{Training}
\label{subsec:training}
We typically let four AIs play 64 games then train for one batch. What's more, inspired by online learning, we will also let neural network review data of last two batches. So the number of data points in each batch is $64\times52\times3=9984$. Then the target function \eqref{eqn::defloss} is optimized by Adam optimizer with $\textrm{lr}=0.001$ and $\beta=(0.3,0.999)$ for 3 iterations.

There is one thing special in our dealing with loss function which deserves some explanations. Normally, it's natural to mask the illegal choice in the probability output of network (i.e. mask them after the softmax layer). However, we mask the output before softmax layer. Or in coding language, we use $\textrm{softmax}(p\times \textrm{legal\_mask})$ rather than $\textrm{legal\_mask}\times\textrm{softmax}(p)$. We find this procedure much better than the other one. A possible explanation is that, if we multiply the mask before softmax, the information of illegal choice is still preserved in the loss, which can help the layer before output to find a more reasonable feature representation.

\subsection{Arena}
\label{subsec:arena}
We provide an arena for different AIs to combat. We define standard protocols for AIs and arena to commute. We set up an server welcoming AIs from all around the world. Every AI obeying our protocol can combat with our AIs and download detailed statistics. We also provide data and standard utility functions for others training usage. More details can be found in our \href{\GongzhuSociety}{GitHub repository}.

\subsection{Details for Evaluation System}
\label{subsec:detail_eval_system}

In section \ref{subsec:eval_system}, we introduced Winning Point over Mr. Greed (WPG). For this evaluation system to be well-defined, WPG should be transitive, or at least approximately transitive, i.e. program with higher WPG should play better. In this section, We will introduce a statistics $\varepsilon$ that measures the intransitivity of an evaluation function, then show that WPG is nearly transitive by numerical experiments. 

We first define $\xi_{ij}$ to be the average winning score of playing strategy $\pi_i$ against strategy $\pi_j$. Then let us start by considering two extreme cases. A game is transitive on $\Pi$, if 
\begin{equation}
    \xi_{ij}+\xi_{jk}=\xi_{ik}~~\forall \pi_i,\pi_j,\pi_k\in \Pi,
\end{equation}
where $\Pi$ is a subspace of the strategy space. As a famous example for intransitive game, 2-player rock-paper-scissor game is not transitive for strategy tuple $(\text{Always play rock},~\text{Always play paper},~\text{Always play scissor})$.

In the middle of the totally transitive and intransitive games, we want to build a function $\varepsilon_{\Pi}$ to describe the transitivity of policy tuple $\Pi=\left(\pi_1,\pi_2,...\pi_n\right)$. To better characterize the intransitivity, the function $\varepsilon$ should have following properties:
\begin{enumerate}
    \item[a)] take value inside $[0,1]$, and equal $0$ when evaluation system is totally transitive, $1$ when totally intransitive;
    \item[b)] be invariant under translation and inflation of scores;
    \item[c)] be invariant under reindexing of $\pi_i$'s in $\Pi$;
    \item[d)] take use of combating results of every triple strategies $(\pi_i,\pi_j,\pi_k)$ in $\Pi$. There are $C^3_n=n(n-1)(n-2)/6$ such triples.
    \item[e)] will not degenerate (i.e. approach to 0 or 1) under infinitely duplication of any $\pi_i\in\Pi$;
    \item[f)] be stable under adding 
    similar level strategies into $\Pi$.
\end{enumerate}
We define $\varepsilon$ to be
\begin{equation}\label{eqn:defoftransitivity}
    \varepsilon_{\Pi}=\frac{
    \sum\limits_{i<j<k}\left|\xi_{ij}+\xi_{jk}-\xi_{ik}\right|^2
    }{
    \sum\limits_{i<j<k}|\xi_{ij}+\xi_{jk}-\xi_{ik}|\left(|\xi_{ij}|+|\xi_{jk}|+|\xi_{ik}|\right)}.
\end{equation}
When there are only three strategies $i,~j$ and $k$, this definition of $\epsilon$ reduces to the form of
\begin{equation}
    \varepsilon_{ijk}=\frac{|\xi_{ij}+\xi_{jk}-\xi_{ik}|}{|\xi_{ij}|+|\xi_{jk}|+|\xi_{ik}|}.
\end{equation}
Equation \eqref{eqn:defoftransitivity} can be seen as average of $\varepsilon_{ijk}$ with weights $|\xi_{ij}+\xi_{jk}-\xi_{ik}|\left(|\xi_{ij}|+|\xi_{jk}|+|\xi_{ik}|\right)$. If one strategy $\pi_{i0}\in \Pi$ is duplicated infinity times, definition \eqref{eqn:defoftransitivity} will approach
\begin{align}
    &\varepsilon_{\Pi\textrm{ with infinite }\pi_{i_0}}=\nonumber\\
    &\frac{
    \sum\limits_{j<k}\left|\xi_{i_0j}+\xi_{jk}-\xi_{i_0k}\right|^2
    }{
    \sum\limits_{j<k}|\xi_{i_0j}+\xi_{jk}-\xi_{i_0k}|\left(|\xi_{i_0j}|+|\xi_{jk}|+|\xi_{i_0k}|\right)}.
\end{align}
In other words, the $|\xi_{ij}+\xi_{jk}-\xi_{ik}|$ factor in weights guarantees property (e). The $\left(|\xi_{ij}|+|\xi_{jk}|+|\xi_{ik}|\right)$ factor guarantees property (f), because, by definition, ``similar level strategies" means $\left(|\xi_{ij}|+|\xi_{jk}|+|\xi_{ik}|\right)$ is small. One can easily verify that this $\varepsilon$ obeys other properties in the list above.

Using the definition in equation \eqref{eqn:defoftransitivity}, we calculate the transitivity of the strategy tuple of 4 different AIs as shown in Table. \ref{tab:transitivity}.
\begin{equation}
\label{eqn:epsilonstatitics}
    \varepsilon_{\{\textrm{R,I,G,ZTS}\}}=0.03\pm0.02
\end{equation}
The $0.02$ in equation above is the standard deviation comes from evaluation of two agents. The statistics in \eqref{eqn:epsilonstatitics} is not significantly from 0. Therefore, playing with Mr.Greed and calculating average winning score is a good choice for evaluating the strength of AI.

\subsection{Examples for Irrelevant Cards}
\label{subsec:example-irrelevant-cards}
Table \ref{tab:irrelevant-card-D4} shows an example for irrelevant cards. Roughly speaking, irrelevant cards are those cards (i) that don't result in large change in both their own correction factors and other cards' correction factors if they are switched to other player, and (ii) whose correction factors don't change much if other cards are switched. Correction factor is defined in equation \eqref{eqn:defofcorfactor}. We can see from table \ref{tab:irrelevant-card-D4} that, with or without D4, the correction factors of other cards are almost the same. Also different combinations of other cards (e.g. ``D7, D8", ``D6, D7, DJ" or ``D10, DK") will not affect the correction factor of D4.

\begin{table}[!ht]
    \centering
    \begin{tabular}{c|cc}
         Cards Guessed& $\gamma$ Without D4& $\gamma$ With D4\\
         \hline
         \rowcolor{green!60!yellow!30} D7& 1.0000& 1.0000\\
         \rowcolor{green!60!yellow!30} D8& 0.9897& 0.9912\\
         \rowcolor{green!60!yellow!30} D4& -     & 0.9188\\
         \hline
         \rowcolor{green!80!yellow!50} D6& 0.9010& 0.9043\\
         \rowcolor{green!80!yellow!50} D7& 1.0000& 1.0000\\
         \rowcolor{green!80!yellow!50} DJ& 0.7398& 0.7321\\
         \rowcolor{green!80!yellow!50} D4& -     & 0.9233\\
         \hline
         \rowcolor{green!60!yellow!30} D10& 1.0000& 1.0000\\
         \rowcolor{green!60!yellow!30} DK& 0.9538& 0.9548\\
         \rowcolor{green!60!yellow!30} D4& -     & 0.9519\\
         \hline
    \end{tabular}
    \caption{An example for irrelevant card. Here D4 is the irrelevant card. $\gamma$ is correction factor of corresponding cards.}
    \label{tab:irrelevant-card-D4}
\end{table}

From figure \ref{fig:val_var}, we know that irrelevant cards are most likely appear in spade. Table \ref{tab:irrelevant-card-SJ} is an example for irrelevant card SJ. Apart from SJ, other small cards, S2 to S7 are also highly likely to be irrelevant cards in this situation.

\begin{table}[!ht]
    \centering
    \begin{tabular}{c|cc}
         Cards Guessed& $\gamma$ Without SJ& $\gamma$ With SJ\\
         \hline
         \rowcolor{green!60!yellow!30} S5& 0.9424& 0.9464\\
         \rowcolor{green!60!yellow!30} S10& 1.0000& 1.0000\\
         \rowcolor{green!60!yellow!30} SJ& -     & 0.9598\\
         \hline
         \rowcolor{green!80!yellow!50} S8& 0.9560& 0.9585\\
         \rowcolor{green!80!yellow!50} S10& 1.0000& 1.0000\\
         \rowcolor{green!80!yellow!50} SQ& 0.4818& 0.4950\\
         \rowcolor{green!80!yellow!50} SJ& -     & 0.9528\\
         \hline
         \rowcolor{green!60!yellow!30} S7& 1.0000& 1.0000\\
         \rowcolor{green!60!yellow!30} SK& 0.8302& 0.8331\\
         \rowcolor{green!60!yellow!30} SJ& -     & 0.9677\\
         \hline
    \end{tabular}
    \caption{An example for irrelevant card. Here SJ is the irrelevant card. $\gamma$ is correction factor of corresponding cards.}
    \label{tab:irrelevant-card-SJ}
\end{table}

\subsection{Select Important History in IEC}
\label{subsec:impotant-history-IEC}
Like what we have discussed in section \ref{subsec:integral-over-EC}, as an attention mechanism and to save computation resources, some ``important" history slices are selected based on statistics in section \ref{sec:empiricalanalysis} in IEC algorithm. From figure \ref{fig:reg} and \ref{fig:val_var}, we can see that, the cards smaller than 8 always have small correction factors in its suit and lower value variance. This means that the history slices with card played smaller than 8 are highly likely to be unimportant. So we only select history slices with the card played greater than 7 in IEC algorithm. Another selection rule is that, when a player is following the suit, history slices of other suits are also not important.

\subsection{Average over unknown Information in IEC}
\label{subsec:blind-info-IEC}
\begin{figure}[!ht]
    \centering
    \includegraphics[width=.45\textwidth]{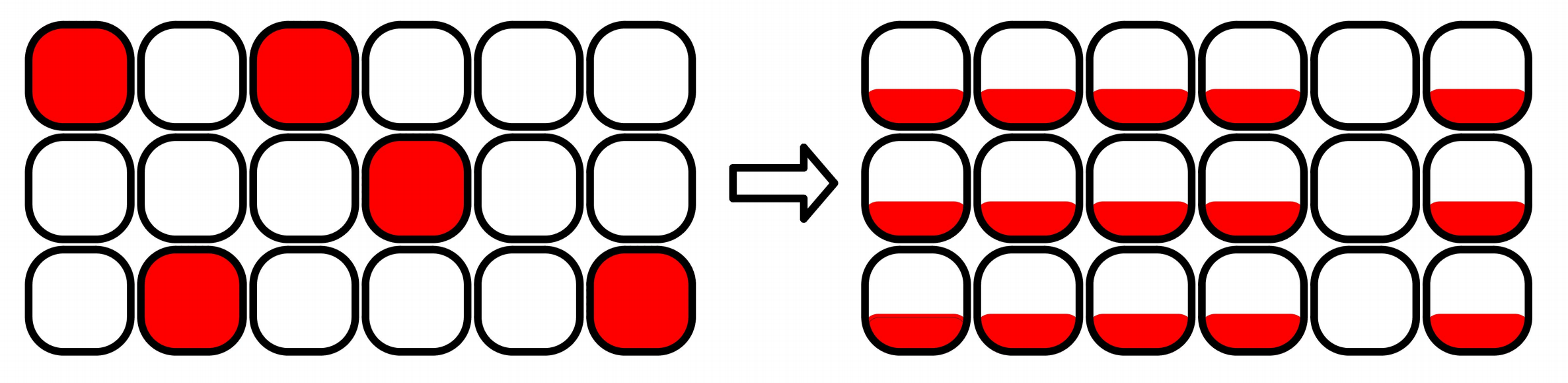}
    \caption{The network input representing other's hands is averaged in IEC.}
    \label{fig:average_input}
\end{figure}

In section \ref{subsec:integral-over-EC}, we introduced IEC algorithm. However, our network's input requests other's hands. We should not give hands of this scenario directly to neural network, because that player cannot know the exact correct hands of other players. As a work around, we average the hands information and give it to neural network, shown in figure \ref{fig:average_input}. In other words, we replace the ``onehot" in input representing other's hands with ``triple-$1/3$-hot".

The replacement from ``one" to ``triple-$1/3$''is not standard. Table \ref{tab:blind-info} shows that the performance will not deteriorate under this nonstandard input.

\begin{table}[!ht]
    \centering
    {\rowcolors{2}{green!80!yellow!50}{green!60!yellow!30}
    \begin{tabular}{c|c}
    Input method& WPG\\
    Standard input (``onehot")& $1\pm16$ \\
    Averaged Input (``triple-$1/3$-hot")& $7\pm4.6$
    \end{tabular}}
    \caption{Raw network performance under standard input and averaged input. Raw network means that we directly use policy network in playing rather than MCTS.}
    \label{tab:blind-info}
\end{table}

\subsection{Proof of Equation \eqref{eqn:int_p_result}}
\label{subsec:detail-IEC}
Once \textit{irrelevant cards} are defined, $c_{-i}$ is divided in to three parts: cards already played, relevant cards and approximately irrelevant cards. Form now on, we will refer approximately irrelevant cards as irrelevant cards. The $J_j$ in equation \eqref{eqn:int_p} is the sum of all of irrelevant cards' correction factors in player $j$, while the $Y_t$ in equation \eqref{eqn:int_p} is the sum of the other cards' correction factors of the player who played stage $t$.
\begin{equation}
    \sum_{\substack{c_k\in\textrm{irrelevant cards}\\ \textrm{in player }j}}\gamma(c_k)\triangleq J_j~,~
    \sum_{\substack{c_k\in\textrm{cards played}\\\cup~\textrm{relevant cards}}}\gamma(c_k)\triangleq Y_t
\end{equation}

Notice that by definition of \textit{irrelevant cards}, $\sum_{j=1}^3 J_j$ should keep unchanged in different scenarios for this decision node, denoted by $J$.
\begin{equation}
    J_1+J_2+J_3=\textrm{Const}\triangleq J.
\end{equation}
The distribution of $J_j$ is polynomial. In most situations, there are always many (5 or more) irrelevant cards. By central limit theorem, a polynomial distribution can be approximated by a multivariate normal distribution. Since $J_3=J-J_1-J_2$, we can derive the marginal distribution of $J_1$ and $J_2$. 
We adopt this approximation and replace the summation over all permutations of irrelevant cards in equation \eqref{eqn:int_p} by an integral
\begin{equation}\label{eqn:int_p_detail}
\begin{aligned}
&p(h^u|c_{-i})\approx \frac{3^N}{2\pi|\Sigma|}\iint_{x_{1,2}>-J/3}^{x_1+x_2<J/3}\\
&\qquad\qquad\quad\text{exp}\left(-\frac{1}{2}(x_1,x_2)\Sigma^{-1}(x_1,x_2)^T\right)\\
&\qquad\qquad\quad\prod_{t=0}^{u-1}\frac{\gamma(a_{t+1},h^t,c_{j(t+1)})}{Y_{t+1}+J/3+x_{j(t+1)}}\,\mathrm{d} x_1\mathrm{d} x_2\\
&=\frac{3^N J^2}{4\pi|\Sigma|} \prod_{t=0}^{u-1}\frac{\gamma(a_{t+1},h^t,c_{j(t+1)})}{Y_{t+1}+J/3}+O(\xi^3)\\
&=3^N \prod_{t=0}^{u-1}\frac{\gamma(a_{t+1},h^t,c_{j(t+1)})}{Y_{t+1}+J/3}+O(\xi^3)
\end{aligned}
\end{equation}
where $N$ is the number of irrelevant cards, $J$ the sum of correction factor of all irrelevant cards, $x_j=J_j-J/3$ and $\xi$ a real number between $0$ and $2/3$. The last equality is because $\Sigma$ satisfies
\begin{equation}
\begin{aligned}
&\frac{1}{2\pi|\Sigma|}\iint_{x_{1,2}>\frac{-J}{3}}^{x_1+x_2<\frac{J}{3}}
e^{-\frac{1}{2}(x_1,x_2)\Sigma^{-1}(x_1,x_2)^T}\,\mathrm{d} x_1\mathrm{d} x_2\\
=&1-O(\xi^3).
\end{aligned}
\end{equation}
The result of equation \eqref{eqn:int_p_detail} is equation \eqref{eqn:int_p_result} in main text.

\end{document}